\documentclass[aps,superscriptaddress]{revtex4}
\usepackage{graphicx}
\usepackage{amsmath}   
\newcommand \eps {\epsilon}

\newcommand \lan {\langle}
\newcommand \ran {\rangle}
\newcommand \be {\begin{equation}}
\newcommand \ee {\end{equation}}

\begin{document}
\title{Temperature shifts in the Sinai model: static and dynamical effects}

\author{Marta Sales} 
\affiliation{Departament de F\'{\i}sica Fonamental, \\
Facultat de F\'{\i}sica, Universitat de Barcelona\\
Diagonal 647, 08028 Barcelona, Spain}

\author{Jean-Philippe Bouchaud }
\affiliation{Service de Physique de
l'\'Etat Condens\'e,
 Centre d'\'etudes de Saclay, Orme des Merisiers, 91191 Gif-sur-Yvette
Cedex, France} 

\author{F\`elix Ritort} 
\affiliation{Departament de F\'{\i}sica Fonamental, \\
Facultat de F\'{\i}sica, Universitat de Barcelona\\
Diagonal 647, 08028 Barcelona, Spain}
\begin{abstract}
We study analytically and numerically the role of temperature shifts in the simplest 
model where the energy landscape is explicitely hierarchical, namely the Sinai
model. This model has both attractive features (there are valleys within valleys 
in a strict self similar sense), but also one important drawback: there is no
phase transition so that the model is, in the large size limit, 
effectively at zero temperature. We compute various static chaos 
indicators, that are found to be trivial in the large size limit, but
exhibit interesting features for finite sizes. Correspondingly, for finite times, 
some interesting rejuvenation effects, related to the self similar nature of the potential,
are observed. Still, the separation of time scales/length scales with 
temperatures in this model is much weaker that in experimental spin-glasses.
\end{abstract}
\maketitle
\newcommand{\eq}[1]{(\ref{#1})}
\noindent
\section{Introduction}

The phenomenology of very different glassy systems (spin glasses,
structural glasses, pinned defects) exhibits striking similarities.
This is often discussed in terms of complex energy landscape,
with some universal features in the organisation of valleys, saddles
and barriers \cite{Gold69,SW82,KL96,I01}. Mean-field models suggest that the energy landscape of
some systems are hierarchical, with a self-similar organisation of
valleys within valleys \cite{MPV}. This picture has been advocated to interpret
\cite{Vincent,BD95,CK99} the striking rejuvenation and memory effects
during temperature cycling experiments, observed first in
spin-glasses, but also in disordered ferromagnets, dipolar glasses,
polymer glasses (PMMA) or gelatin \cite{HV00,JV98,VH96,Parker}.

Another general framework to understand disordered systems is the
droplet picture, developed in the context of spin-glasses and of
pinned manifolds (domain walls, vortices, etc.)  \cite{FH88,BM87}.
The approach aims at describing the physics in terms of localized
excitations of different size (droplets), with an energy scale that
grows with their size. As emphasized in \cite{B99,BD01}, these two
approaches are in fact complementary, and the dynamics in the droplet
model is naturally hierarchical, due to the strong separation of time
scales with length scales and with temperature. Temperature plays the
role of a microscope since the active length scales that contribute to
the dynamics are very different at different temperatures.

In both pictures, memory conservation is due to this separation of
time scales with temperature \cite{YL01, BB02} (see also \cite{BH02}). The origin of {\it rejuvenation},
however, is different. In the hierarchical landscape framework, a
small temperature drop (from $T_1$ to $T_2=T_1-\Delta T$) reveals
finer details of the random energy landscape, among which the system
must reequilibrate. In other words, small length scales, that are
`unpinned' at temperature $T_1$, freeze at $T_2$, thereby producing a
strong out of equilibrium signal (rejuvenation).  This scenario can be
given some precise meaning in the context of the Random Energy Model
({\sc rem}) close to its critical temperature \cite{SB01}, or in the
generalized (multilevel) {\sc rem} where rejuvenation and memory
effects very similar to experiments can indeed be observed
\cite{SN00,K01}. On the other hand, rejuvenation effects in the
droplet model have been attributed to `temperature chaos': for any
temperature difference $\Delta T$, the equilibrium configurations at
the two temperatures are completely different beyond a certain length
scale $\ell_{\Delta T}$, called the overlap length, that diverges when
$\Delta T \to 0$ as a power-law. Correspondingly, the correlation of
the free-energies at the two temperatures are thought to decay with
the size of the system $L$ as $\exp(-L/\ell_{\Delta T})$. This
scenario, postulated for spin-glasses in \cite{BM87,FH88}, has been
given credit recently in the context of directed polymers
\cite{FH91,SY02,dSB02}. In this case, both analytical arguments and
numerical results point towards the existence of an overlap length,
although the free-energy decorrelation appears to be much slower than
exponential in the size of the system. For spin-glasses, the status of
temperature chaos is still controversial and unclear as compared to
the influence of other, stronger perturbations such as magnetic field
changes or coupling strength changes. Results in mean-field spin glasses 
\cite{KONDOR89,R01} and numerical calculations
for small systems \cite{RITORT94} hint at extremely weak chaotic
effects in temperature. This conclusion extends also to short-range
systems where chaotic effects in the equilibrium properties seem to be
extremely weak \cite{NY97,NIFLE98}. Chaotic effects, if any, 
seem to appear only for quite large systems
even when $\Delta T/T$ is of order one \cite{BM00,SM02}. 
Nevertheless, some recent spin-glass experiments
were interpreted in terms of an overlap length \cite{YJ02}, whereas
numerical simulations of the 4D Edwards-Anderson model has revealed
rejuvenation due to small scale freezing without chaos
\cite{BB02}. (For a detailed study
in the 3D Edwards-Anderson model and a discussion about why
rejuvenation effects are not observed in simulations see \cite{PR01,BB02}).

In view of the controversy, we feel that more work on the subject is
needed. The aim of the present paper is to study in details the
question of `temperature chaos' on the Sinai model, which is the
simplest model where the energy landscape is {\it explicitely
hierarchical}, and for which many exact results are available
\cite{BC90,BG90,LD98,DM99,MD02}. The Sinai model was discussed in this
context in \cite{VH96,B99}; preliminary results were obtained by
H. Yoshino, but no systematic study had been performed.

The Sinai model is an example of a one dimensional self-similar
potential with long-range correlations. In high dimensions, this
problem is equivalent to a mean-field spin-glass with a continuous
Replica Symmetry Breaking solution \cite{MP91,ClD96}. In one
dimension, however, there is no phase transition: the long time, large
scale behaviour of the system is ruled by the zero-temperature fixed
point, where the deepest minimum dominates. This is because the
entropy in this model is of order one, whereas the energy scale grows
as $\sqrt{L}$. Since the location and energy of the minimum is
temperature independent, there cannot be any true temperature chaos in
this problem. However, there are interesting static and dynamical
crossover effects, as a function of the size of the system or of the
time, that we study in details in this paper. We find that
rejuvenation and memory effects are present, {\it in embryo}, in the
Sinai model.

The paper is organized as follows. In the next section we present the Sinai model. 
In section III we analyze the statics of the model. In section IV we discuss the dynamics, 
first by studying aging in the correlation length and then by studying the a.c. susceptibility 
during temperature cycles. In section V we give the conclusion derived from our analysis.

\section{The model: Sinai potential} 

The Sinai model belongs to a wider class of random potential models. It 
describes the dynamics of a point particle under the action of a random, uncorrelated 
force field 
$F(x)$ which models several physical situations \cite{BG90}, such as the motion of 
domain wall in the
Random Field Ising Model \cite{CC02}, or the motion of a dislocation kink. More recently, this 
model was argued
to be relevant to describe some tapping experiments in sandpiles \cite{Lud01} and the unzipping transition of DNA \cite{LN01}. Interestingly, 
the effective potential 
acting on the end point of the disordered 
directed polymer in 1+1 dimensions is also of the Sinai type \cite{Mezard,Parisi,BO}. 
However, in this
case, the effective potential itself becomes temperature dependent, and the role of 
temperature changes in the directed polymer is much more subtle \cite{FH91,SY02,dSB02}. 

In the following, we will study  the discrete Sinai model where each
site corresponds to a different state or configuration with energy
$V_i$. The system consists of a box of length $L$ in which we generate
a random potential. Each sample of the random potential is constructed
as follows: at each site $i \in [1,L]$, we generate a random Gaussian
force $f_i$ with zero mean and variance $\overline{f_i
f_j}\;=\;\sigma^2\;\delta_{i,j}$. All along this paper, 
we have taken $\sigma=1$ in the numerical simulations.
The potential in each site is the sum of the forces in the previous
sites $V_i=-\sum_{j=1,i}\; F_j$, and is thus a {\it random walk} as a
function of the position. Thus correlations (and equivalently
barriers) grow like:
\begin{equation}
\overline{(V_i-V_j)^2}=\sigma^2 \; |i-j|. 
\label{eq5}
\end{equation}
This model has no thermodynamic transition so that in the limit $L\to \infty$, the 
physics is dominated by the $T=0$ glassy fixed point \cite{OM93,CM98}. As is obvious from Eq.(\ref{eq5}), 
a rescaling of the length by a factor $b$ is equivalent to a change of the scale of the potential by a factor
$\sqrt{b}$, or of the temperature by a factor $1/\sqrt{b}$.
As far as the statics are concerned, this means that being at low temperature in a small 
system is equivalent to having a larger system but a higher temperature. For thedynamics 
a change in temperature leads to both a change in length scales and timescales.

\section{Static Chaos}

\subsection{Observables}
Our aim is to investigate how the thermodynamic properties of the Sinai model change when 
comparing the same system at two different temperatures. 
In order to probe the change in the free-energy landscape we have 
studied two different quantities, (i) the correlation 
function of the free-energy fluctuations at different temperatures and (ii) 
the correlation of the particle position. More precisely, we have studied the following observables:    
 \begin{itemize}

\item{Free-energy correlations:} It consists in measuring the free energy fluctuations 
of the system at different temperatures averaged over the disorder:
\begin{equation}
C_F(L,T_1,T_2)=\frac{\overline{\Phi_{T_1}\Phi_{T_2}}}
	{\sqrt{\overline{\Phi_{T_1}^2}}\sqrt{\overline{\Phi_{T_2}^2}}} \qquad 
	\Phi_{T}\equiv F_T-\overline{F_{T}},
\label{eq1}
\end{equation}
where $F_T$ is the free energy at temperature $T$ and $\overline{(..)}$ stands for the average over different realizations of the potential.
This quantity was originally proposed by Fisher and Huse \cite{FH91} 
to study temperature chaos in the directed polymer problem. The fact that this correlation 
tends to zero with the size of the system shows that different energy 
valleys contribute to the total free-energy at different temperatures. More recently this quantity has 
been used to study the chaotic properties in the {\sc rem} \cite{SB01}, 
and in the directed polymer problem when different type of perturbations are introduced \cite{SY02}, or for
directed polymers on a hierarchical lattice \cite{dSB02}.

\item{Fluctuations of position:} A more geometrical way to visualize `temperature chaos' 
which could be of direct interest in some cases, for example the 
zipping and unzipping problem of DNA \cite{LN01}, is to consider the following quantity.
To each site $i$, we associate a position $x_i=i/L$ and compute:
\begin{equation}
d_{1,2} \equiv \overline{(\lan x\ran_{T_1}-\lan x\ran_{T_2})^2},
\label{eq2}
\end{equation}
where $\lan .. \ran_T$ is the thermal average at temperature $T$,  
and $\overline{(...)}$ is the disorder average.

The study of the distance between average positions corresponding 
to different temperatures gives us an indication of the distance between states 
contributing to the partition function ${\cal Z}$ at different temperatures. If typical states  
contributing to ${\cal Z}$ at $T_1$ and $T_2$ are completely different, then 
$d_{1,2}$ 
will remain finite as $L \to \infty$. Note that $d_{1,2}$ has an upper bound 
$d_{1,2} 
\leq {1}/{6}$,
where $1/6$ is the value reached if the occupied sites are completely uncorrelated.

\end{itemize}

Both quantities $C_F$ and $d$ have been studied in the {\sc rem} \cite{SB01}
to show that even if the energy landscape is fixed (in the sense that
there is no reshuffling associated with the different valleys as
temperature is changed), the model exhibits temperature chaos
around the critical temperature, where the Boltzmann weight
`condensates' into a finite number of sites. Because there is no
finite temperature transition in the Sinai model, one only observes
mild effects under a temperature change, that are maximum around the
crossover temperature $T \sim L^{1/2}$ (see below).

Numerically we have investigated how these quantities behaves as a function of $L$ 
for different temperature changes.  
We have averaged over 2000 potential samples of sizes ranging from $L=1$ to $L=2^{14}$.

\subsection{Thermodynamics of the model}
\label{t}

The thermodynamics of the Sinai model has been well studied
\cite{OM93,BK,CM98}. In the large $L$ limit temperature is
irrelevant. The system is frozen or localized in the minima of the
potential, so that physical observables are governed by the ground
state and its fluctuations. The free-energy grows as $\overline{F}\sim
- \sqrt{L}$ independently of temperature and the entropy reaches an
$L$-independent value which depends on temperature
$\overline{S}\;=\;A+2\ln T$. The free-energy fluctuations are also
essentially dominated by the fluctuations of the deepest valley at any
temperature. Therefore, it is reasonable to expect that in the
thermodynamic limit a change in temperature has no significant effect
on physical observables since the ground state properties are
temperature independent.\\

The finite size corrections to the thermodynamic behaviour enter through the variable 
$g=\sigma \sqrt{L}/T$. Corrections are important when $g\sim 1$, or $L \sim 
L^*=(T/\sigma)^2$, signaling the crossover between two different limits:

\begin{itemize}
\item{$L \ll L^*$:} energy differences are much smaller than the temperature, so that all 
the sites contribute to the partition function.
\item{$L\gg L^*$:} energy differences and barriers are huge and the particle is localized; 
only a few sites within the deepest valley contribute significantly to the partition function.    
\end{itemize}
This crossover is clearly observed in the behavior of the fluctuations of the mean position
(in units of $L$):
\begin{equation}\Delta^2 = \overline{\lan x\ran^2}-\overline{\lan x\ran}^2.
\end{equation}

\begin{figure}[h]
\begin{center}
\includegraphics*[scale=0.4]{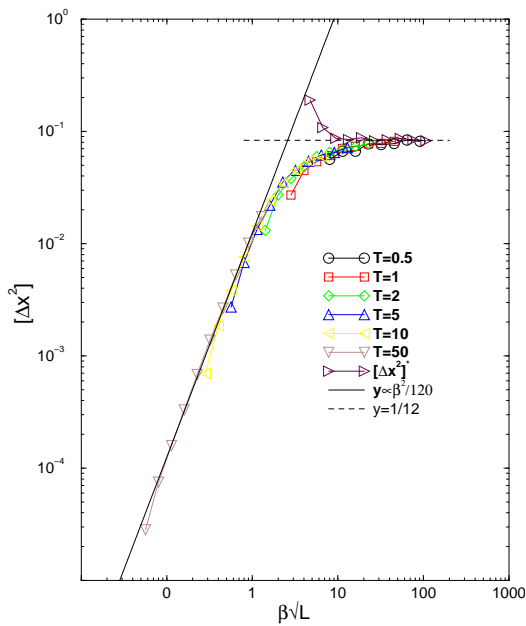}

\vskip 0.05in
\caption{Sinai Potential: $\Delta^2$ plotted versus the scaling variable $g=\sigma 
\sqrt{L}/T$ for $T=1,\;2,\;10,\;50\;$and$\;100$. The dashed line corresponds to 
the position fluctuations at zero temperature. For small $g$ we have plotted the 
high temperature prediction ${g^2}/{120}$. 
\label{dxx}}
\end{center}
\end{figure}

In figure \ref{dxx} we show the results for different temperatures versus the scaling variable 
$g$ defined above. The small $g$ behavior can be computed using a high-temperature expansion
of the partition function: 
\begin{equation}\Delta^2 =\frac{g^2}{120} +{\cal O}(g^4).
\label{eq6}
\end{equation} 
As we can see in the figure, the behavior for small $g$ matches nicely this prediction.
For large $g$ the system is completely governed by the $T=0$ behavior, 
thus average position fluctuations should approach the fluctuations of the ground state 
position. In figure \ref{dxx} we also plot 
the ground-state position fluctuations as a function of $\sqrt{L}$, that converges 
towards $1/12$. 
Numerically we can see that this crossover takes place at $g\simeq\;3$, corresponding
to $L^*\simeq\;10 (T/\sigma)^2$     

This static crossover can be mapped to the dynamical crossover 
from a Brownian diffusive regime to an activated regime as we will see 
in the analysis of the correlation length in section \ref{lcorr}.

\subsubsection{Free-energy fluctuations}
\label{fe}
As we can see from the inset in figure \ref{cf}, the free-energy correlations are a 
non-monotonous function of $L$ ~\footnote{We have numerically checked that the
energy-correlations have the same behavior}. For small $L$ the free-energy at different 
temperatures start decorrelating but at a certain length, $L_F(T_1,T_2)$, $C_F$ reaches 
a minimum. Then, it starts increasing back and reaches $C_F=1$ when $L \to \infty$ 
regardless of the temperature difference (provided it is finite) as expected. One
can distinguish between two different regimes:
\begin{itemize}
\item{$L \ll L^*_F$}\\
For small system sizes/large enough temperatures the energy landscape is essentially flat, 
so that all the sites contribute to ${\cal Z}$. Therefore, in this regime 
a high temperature expansion is expected to yield the correct behavior of $C_F$. 
The result is the following:
\begin{equation}1-C_F \;\propto\;  (\beta_1-\beta_2)^2 \sigma^2 L 
\qquad(\Delta \beta \sigma \sqrt{L}\to 0),
\end{equation}
where $\beta \equiv 1/T$.
\item{$L\gg L^*_{F}$}\\
In this limit the system is governed by the ground state and its fluctuations, 
thus we expect that for $L \to \infty$ the correlation is perfect. However when 
$g_{1,2}=\sigma \sqrt{L}/{T_{1,2}}$ is large but finite, the free energy decorrelates slightly. 
The behavior in the large $L$ limit can be understood by very simple arguments.
Suppose that $T=(T_1+T_2)/2$ is small, and that the relative temperature change is also small: 
$\eps =(T_1-T_2)/{T} \ll 1$. Using $\partial F/\partial T = - S(T)$, 
where $S(T)$ is the entropy at temperature $T$, one finds:
\begin{equation}\Delta F= F(T_2)-F(T_1)= \Delta T \; S(T)+\;{\cal O}(\eps^3), \qquad \Delta T=T_1-T_2.
\label{eq10}
\end{equation}
Note that this relation is also true for the fluctuating part $\Phi$ of $F$.
Substituting \eq{eq10} into the expression for $C_F$ and expanding for small $\eps$, we 
finally get:
\begin{equation}C_F\approx 1- \Delta T^2 \frac{\overline{S^2}-\overline{S}^2}{2\overline{\Phi^2}}. 
\label{eq15}.
\end{equation}
We show in Appendix A that the entropy fluctuations tend to a numerical constant $K$
that we compute in the large $L$ in limit. 
Thus recalling that free-energy fluctuations scale as $\overline{\Phi^2}=K' \sigma^2 L$, where
$K'$ can be computed from \cite{BK}, we finally find the following scaling behavior for $1-C_F$:
\begin{equation}1-C_F\;~~\stackrel{\sim}{{L\to \infty}} \;\frac{K \Delta T^2}{K'\sigma^2 L}.
\label{eq16} 
\end{equation} 

\item{Crossover}\\ 

The crossover between both regimes will take place at a certain length 
$L^*_F$ such that $\Delta \beta \sigma \sqrt{L^*_F}\sim \Delta T/\sigma \sqrt{L^*_F}$, 
which yields $L_F^*\sim T_1 T_2/\sigma^2$. The maximum of $1-C_F$ actually occurs for 
$L^*_F\approx 10\; T_1T_2/\sigma^2$. 
This suggests the following scaling form
for $1-C_F$:
\begin{equation}1-C_F = \frac{\Delta T^2}{\sigma^2 L^*_F}f\left(\frac{L}{L^*_F}\right),
\end{equation}
with $f(x) \sim x$ for $x \to 0$ and $f(x) \sim 1/x$ for $x \to \infty$. 
The resulting scaling plot is shown in figure \ref{cf}. This scaling is acceptable
only when temperature differences are not too large, as one would expect.
 \begin{figure}[h]
\begin{center}
\rotatebox{270}{
\includegraphics*[scale=0.4]{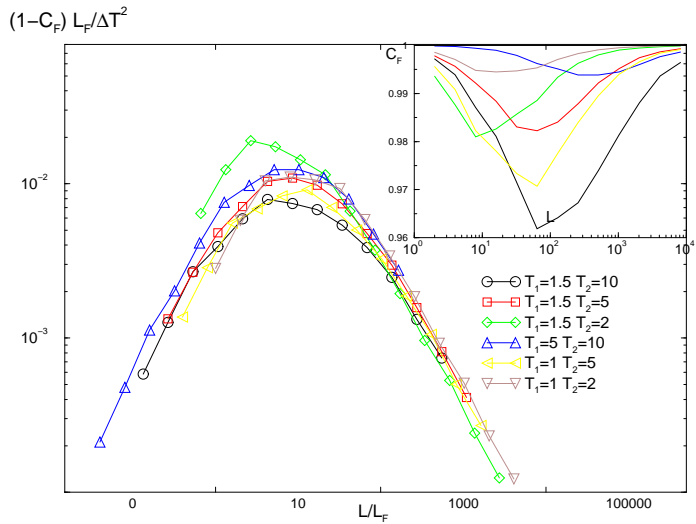}
}
\vskip 0.05in
\caption{Scaling function for the free-energy correlations. 
$(1-C_F)\times \frac{\sigma^2 L^*_F}{\Delta T^2}$ as a function of 
$\frac{L}{L^*_F}$ for different pairs of temperatures ($T_1,T_2$).
\label{cf}}
\end{center}
\end{figure}
\end{itemize}

\subsubsection{Position fluctuations}
\begin{figure}[h]
\begin{center}
\rotatebox{270}{
\includegraphics*[scale=.4]{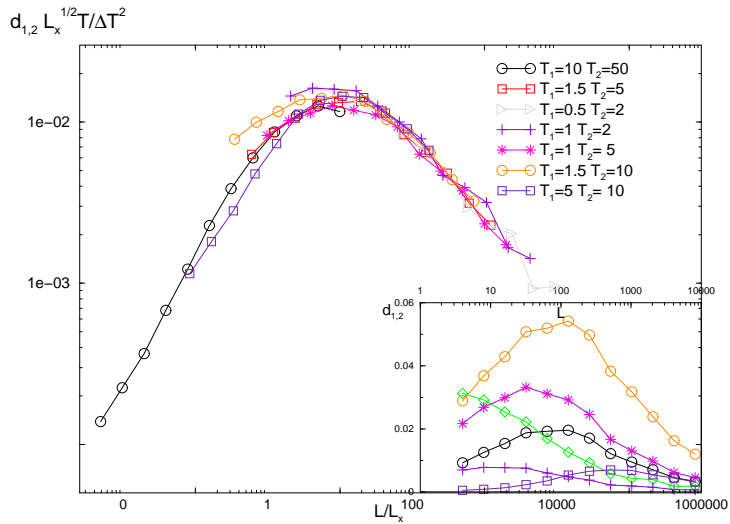}
}
\vskip 0.05in
\caption{Rescaled average distance shift $d_{1,2}\times 
\frac{(T_1T_2 T)^\frac{2}{3}}{(\Delta T)^2}$  plotted versus $L/L^*_d$ for different 
$T_1$ and $T_2$.      
\label{dxs}}
\end{center}
\end{figure}

In figure \ref{dxs} we show the curves corresponding to the average distance shift
$d_{1,2}$ defined by Eq.\eq{eq2} above, for several temperature differences. 
Again we find that this quantity is a non monotonic function of $L$. As a matter of fact, 
the behavior of both $d_{1,2}$ and $C_F$ is similar, and the same type of argument can be
used for both quantities.  
\begin{itemize} 
\item  $L\ll L^*_d$\\
In figure \ref{dxs} we can see that for large temperatures, $d_{1,2}$ increases with 
system size. This behavior can be explained by looking at a high temperature expansion.
As shown in Appendix B, we find in this regime  $d_{1,2}\approx \frac{1}{120}(\Delta \beta)^2 
\sigma^2 L$.
\item $L\gg L^*_d$\\ 
In the large $L$ region the behavior of $d_{1,2}$ should be governed by the 
ground state fluctuations. One starts from the relation:
\begin{equation}
\lan x\ran_{T_1}-\lan x\ran_{T_2}=
\frac{\eps}{T}\left(\lan x\;V_x\ran_T-\lan x\ran_T\lan V_x\ran_T\right), \qquad 
\eps=\frac{\Delta T}{T},
\label{eq19b}
\end{equation}
where $\lan ... \ran_T$ is the thermal average at temperature
$T$. This relation is derived from the fluctuation-dissipation
relation (${\partial \lan x\ran}/{\partial \beta}=-\lan xV_x\ran_c$) and integrating it assuming $\epsilon\ll 1$.
The main 
contribution to $d_{1,2}$ for small temperatures will come from two nearly degenerate valleys
that are a distance $\sim L$ apart. Restricting the thermal averages to these two
valleys, one finds:
\begin{equation}
\lan x\ran_{T_2} - \lan x\ran_{T_1}\approx 
\frac{\eps}{T}{D}\;\sigma \eta\;\sqrt{L}\;
e^{-\beta\sigma \eta \sqrt{L}}, 
\label{eq20}
\end{equation}
where $D$ is the distance between the two valleys, and $\sigma \eta \sqrt{L}$ is their 
energy difference, such that $\eta$ is a positive random variable of order unity.
In order to obtain $d_{1,2}$ we have to compute 
$\overline{(\lan x_2\ran -\lan x_1\ran)^2 }$. This average depends on the joint 
probability distribution of the excitations $P(\eta,D)$. It is reasonable to assume 
that the probability distribution factorizes as $P(\eta,D)=\;h(\eta)\; p(D)$. 
Therefore in the limit $\beta \sigma \sqrt{L} \to\infty$ we have:
 \begin{eqnarray}
\overline{D^2 \eta^2 \, e^{-2\beta \eta \sqrt{L}}}=
 \overline{D^2} \int_0^\infty\; d\eta\; h(\eta)e^{-2\beta \sigma \eta \sqrt{L}}
 \eta^2 {\sim}\;\overline{D^2} 
 \frac{h(0)}{(\beta \sigma)^3L^{3/2}}. 
\label{eq21}
\end{eqnarray}
Recalling that $x$ is the rescaled distance $i/L$, one expects $\overline{D^2} \sim 1$.
Provided $h(0)\neq 0$, one finally finds:
\begin{equation}d_{1,2}\sim \frac{\Delta T^2}{T\sigma\sqrt{L}} \qquad (\beta \sigma \sqrt{L}\to \infty).
\label{eq22}
\end{equation} 

\item{Crossover}\\
As for the free-energy correlation, we can extract a crossover length scale which 
separates both regimes. The crossover length that is obtained is 
$L^*_d\sim(T_1\;T_2)^\frac{4}{3}T^{-\frac{2}{3}}$, that coincides with $L^*_F$ in the
limit $T_1=T_2$. Again, we can try a scaling formula:
\begin{equation}
d_{12}=\frac{(\Delta T)^2}{(T_1T_2 T)^\frac{2}{3}}g\left(\frac{L}{L^*_d}\right)
\qquad{\rm where}~\left\{\begin{array}{cc}
g(x)\sim x &x\to 0\\
g(x)\sim \frac{1}{\sqrt{x}} &x\to \infty
\end{array}\right.
\label{eq23}
\end{equation}
Note that the functions $f$ and $g$ are quite different. This reflects the fact that 
that the two observables probe different mechanisms. In figure \ref{dxs} we show the  
scaling plot for different pairs $(T_1,T_2)$. Notice that {\em all} the curves 
display the maximum at the same value of the scaling variable $\frac{L}{L^*_d}\simeq 10$.
\end{itemize}

\subsection{Discussion}

The outcome of the numerical analysis of the Sinai potential is clear: 
in the thermodynamic limit statistical properties are governed by the 
$T=0$ fixed point. This means that regardless of temperature, for large 
enough system sizes, the system only sees the free-energy valley associated 
to the global minimum. Effectively this situation is equivalent to say that in this 
limit there is no chaos in temperature since the statistical properties are 
those of the minimum of the potential.\\
This situation is very different from what happens in the directed polymer problem. 
This model, as well as the Sinai model, has no thermodynamic transition. 
It only displays a low-temperature glassy phase. In a recent paper \cite{SY02} 
this model has been shown to be extremely sensitive to any perturbation leading 
to the vanishing of correlations between systems at different temperatures (see also
\cite{dSB02,SM02} for a related discussion.)
In this model this is due to existence of anomalous large excitations which have a 
very low free-energy cost. These excitations cost a lot of energy $\Delta E\sim L^{1/2}$
but are very favored entropically, since $T\;\Delta S\sim L^{1/2}$, so that these 
two contributions may cancel to yield a low cost in free-energy. In the Sinai model these 
anomalous excitations do not exist because the entropy is very small (not extensive) 
and never cancels the energy cost ($\sim L^{1/2}$) of such excitations. 

It is interesting to compare the above crossover lengths $L^*_{d,F}$ with the
crossover length at each temperature $L_T^*\sim (T/\sigma)^2$. 
The maximum decorrelation takes place for a system size such that $L^*_2<L=L^*_{d,F}<L^*_1$. 
In this case the system at $T_2$ is already localized whereas the system at 
$T_1$ is still delocalized. The strong influence of temperature shift in this
case is a smeared out version of the infinite susceptibility found in the Random Energy 
Model \cite{SB01}. In this case there is a true finite temperature phase transition, 
and not a mere crossover as in the Sinai case. For larger system sizes, when 
$L^*_1<L^*_2<L$, both systems are governed by the zero temperature fixed point and 
correlations increase.

\section{Dynamics}

The dynamics of this model have been well studied, both analytically \cite{BC90,DM99} and 
numerically \cite{LD98}. Single-time as well as two-time quantities have been 
analyzed. Here, we study one observable of both types; we define in particular an `a.c. 
susceptibility' that should be closely related to the analogous observable studied 
in spin-glasses.

In our simulations we have used boxes of length $L=1024$ with periodic boundary conditions. 
The dynamics has been simulated by the Monte Carlo method using Metropolis algorithm.
For each realisation of the random potential, in order to sample adequately the 
energy landscape we have considered all possible initial conditions ($L=1024$) and we
have averaged over $n \sim 100$ different histories for each starting point. 
The total number of samples used in temperature cycling experiments is around $200-300$.
We have analyzed two different quantities:
\begin{itemize}
\item the correlation (or explored) length:
\begin{equation}\xi^2(t)= \overline{\lan (x(t)- x(0))^2\ran}.
\label{eq25}\end{equation}
The brackets and overline mean that we average over both the $L\times n$ histories and samples respectively.
The initial condition is a uniform distribution equivalent to a quench from infinite 
temperatures. This correlation length only gives information about the `large scale' 
mechanisms and the temperature cycling effects on this quantity can be fully explained in terms of 
effective times. 
\item In order to probe `smaller' length scales that are more sensitive to 
temperature changes, we have defined the following `a.c.' susceptibility $\chi(\omega,t_w)$:
\begin{equation}\chi(\omega,t_w)\equiv
\overline{\left.\left\lan \left(x(t_w+\frac{1}{\omega})-x(t_w)\right)^2\right\ran\right
|_{{\cal P}(x,t_w)}}~~,
\label{eq24}
\end{equation}
where the average is taken over the probability ${\cal P}(x,t_w)$ 
that a particle is at position $x$ at time $t_w$,
with a uniform distribution of particles at time $t=0$ \footnote{We have also
investigated the case where the initial distribution is localized on an 
arbitrary point, with similar results.}. In other words, we measure the
typical extra distance travelled by particles during a time $1/\omega$, weighted by the 
dynamical distribution at time $t_w$. Such a quantity was also considered in \cite{LD98,DM99}.
\end{itemize}

Our main interest in this section is to study the temperature cycling experiments 
which are carried out in spin-glasses that have shown striking rejuvenation and 
memory effects \cite{B99,BD01}. Our main goal here is see to what extent these effects are already 
present in the Sinai model. We have performed numerically the standard temperature 
cycling experiment: quench from infinite temperature down to $T_1$ and let the system 
relax during $t_{w_1}$; then change the temperature to $T_2=T_1+\Delta T$ 
and let the system evolve during $t_{w_2}$ and finally go back to $T_1$. 
We have studied cycles with positive and negative $\Delta T$ for several waiting times 
and frequencies. 

\subsection{The correlation length}
\label{lcorr}
\begin{figure}[h]
\begin{center}
\rotatebox{270}{
\includegraphics*[width=5cm,height=6.5cm]{x.eps}
}
\rotatebox{270}{
\includegraphics*[width=5.25cm,height=6.5cm]{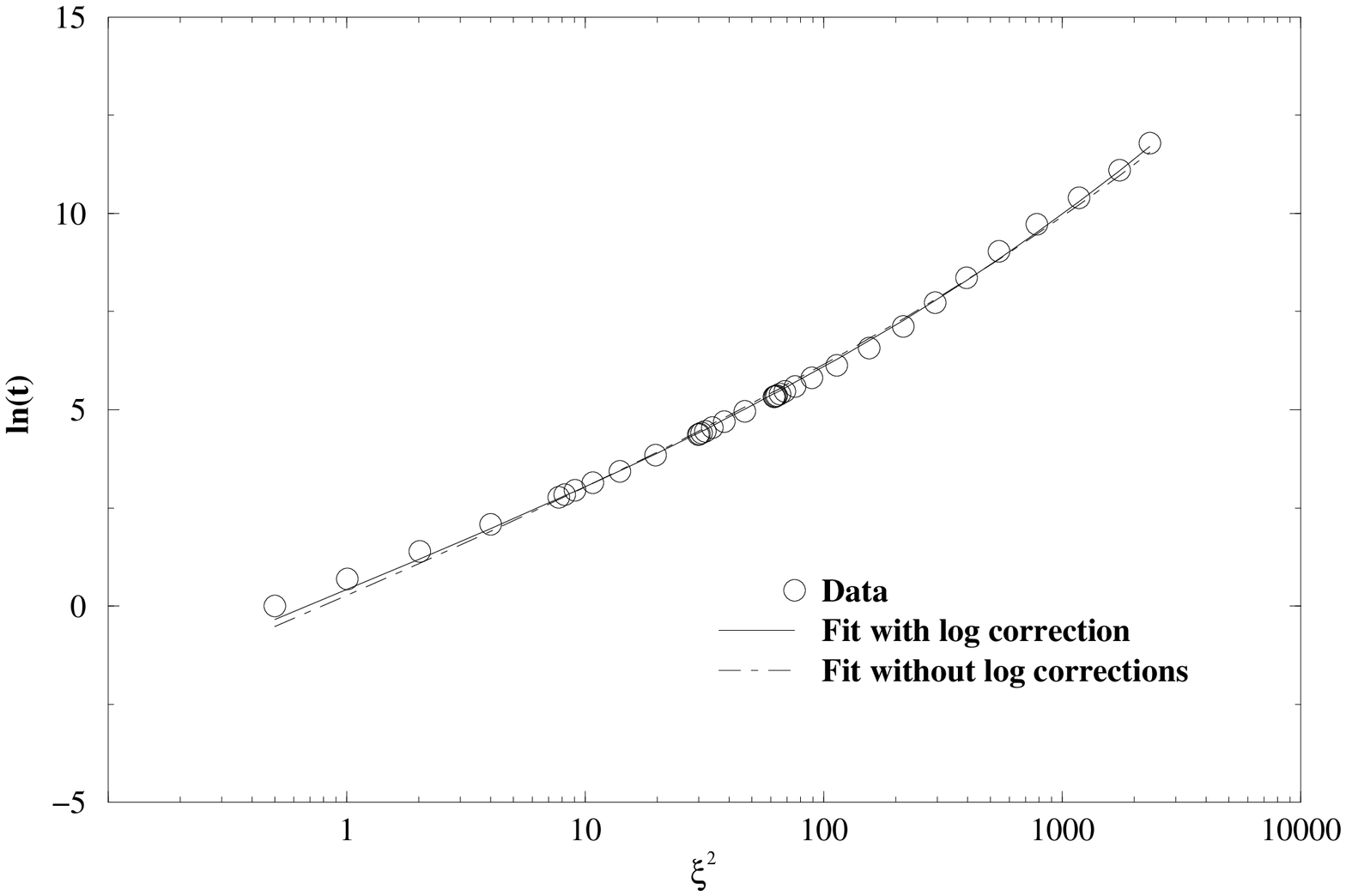}
}
\end{center}
\vskip 0.05in
\caption{{\em Left.} Correlation length versus time for different temperatures. 
The solid line corresponds to the short time diffusive behavior $\xi(t)\sim \sqrt{t}$. 
In the inset we plot ${\xi^2}/{T^4}$ versus the scaling variable $\tau={t}/{200\;T^4}$. 
Averages over $75\times 1024$ histories. {\em Right.} $\ln t$ versus $\xi^2$ for 
$T=0.7$ fitted with the effective function in \eq{eq28b}, with and without the $\ln \ln \ell$ 
correction.} 
\label{x}
\end{figure}

The time evolution for the correlation length at different temperatures is shown in the 
left panel in figure \ref{x}. The growth of the correlation length depends exclusively 
on a temperature dependent microscopic timescale $\tau_0(T)$. This timescale is related 
to the crossover between two different dynamical regimes \cite{BC90,DM99}:\\
\begin{itemize}
\item $t \ll \tau_0$: Short time dynamics where no barriers are present, so that 
in this regime we have usual Brownian diffusion  $\xi^2(t);= \;D\;t$. 
\item $t \gg \tau_0$: Long time, activated dynamics, with an activation time which follows 
an Arrhenius law with a typical barrier $B\;\sim\;\sigma\sqrt{L}$ ~\cite{BC90}, 
leading to $\xi^2(t) \;\sim \;(T/\sigma)^4 \ln^4(\frac{t}{\tau_0})$.
\end{itemize}
The crossover takes place when barriers become comparable to temperature so that 
activation between valleys dominates the dynamics. This crossover is directly related to 
the static crossover found for position or free energy correlations 
from a high temperature regime (no barriers) to a (thermodynamic) low temperature regime 
(see section \ref{t}). The microscopic timescale can be thus identified with the time 
that typically the system takes to explore this static crossover length scale 
$L^*\simeq 100\; (T/\sigma)^2$. Therefore, for $\sigma=1$, $\tau_0(T)=L^{*2}/D \simeq 200\; T^4$, 
($D$ can be evaluated to be $D \approx 0.5$).\\
One therefore expects that the correlation length at a given temperature (\ref{eq25}) 
can be expressed as  $T^4 f(\tau)$, where $f$ is a function of the scaling variable 
$\tau={t}/{\tau_0}$, with:
\begin{equation}\left\{
\begin{array}{ccc}
\tau\ll 1& f(\tau)\sim \tau\\
\tau\gg \tau_0(T)& f(\tau)\sim \ln^4\tau
\end{array}\right.
\label{eq28}
\end{equation}
This scaling behavior works very well, as shown in the inset of Figure \ref{x}, 
where we rescale together all temperatures. Such a crossover, between a short time
growth law and a long time activated behaviour is also expected in the droplet 
description of spin-glasses \cite{FH88,BD01,hajime,BB02}, or directed polymers, where barriers 
grow with the size $\ell$ of the excitations as $B(\ell)=\Upsilon \ell^\psi$, 
where $\Upsilon$ is a function of temperature.  This leads to a logarithmic growth of 
the size of the droplets, $\ell \sim (\ln \frac{t}{\tau_0})^\frac{1}{\psi}$, where 
$\tau_0$ is a microscopic attempt time, possibly renormalized by critical fluctuations \cite{BD01,D01,YN01}.
The Sinai case corresponds to $\psi=1/2$ and $\tau_0 \propto \ell^2$. As a more precise 
description of the crossover, we have shown in Fig. \ref{x} the following fit:
\begin{equation}
t(\ell)=A \;\ell^2\;\exp(B\;\sqrt{\ell \ln \ln \ell}),
\label{eq28b}
\end{equation}
where the $\ln \ln \ell$ accounts for the famous Khinchin iterated logarithm law for 
the maximum of a random walk \cite{Feller}. Note that the
two limiting behaviors for $\xi^2$ in \eq{eq28} consistent with this
last expression. In figure \ref{x} the right 
plot shows $\ln t$ versus $\xi^2$ for $T=0.7$ which are nicely fitted
by the expression above. 

\begin{figure}[h]
\begin{center}
\rotatebox{270}{
\includegraphics[scale=0.4]{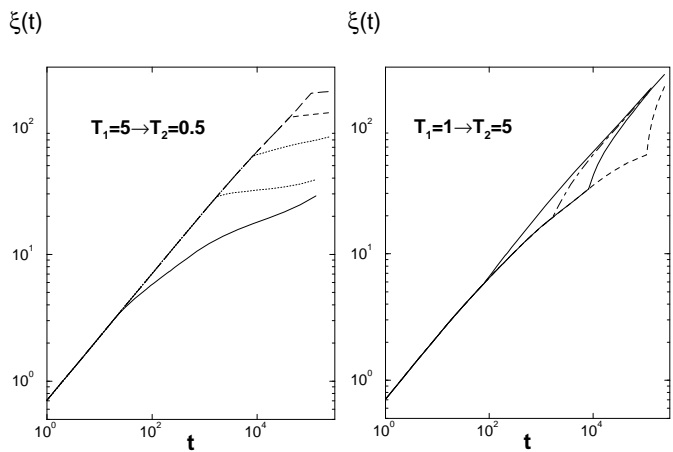}
}
\vskip 0.05in
\caption{Correlation length $\xi(t_{w})$ versus time for the following experiments: {\em Left}) 
spend $t_w$ at $T_1=5$ and then quench the system down to $T_2=.5$ 
for $t_w=\;16,\;1662,\;8192,\;40959$ and $106494$ from bottom to top; 
{\em Right}) spend $t_w$ at $T_1=1$ and then heat the system up to $T_2=5$ for $t_w=\;79,\;1662,\;8192$ and $106494$.
\label{cicle}}
\end{center}
\end{figure}

The correlation length is a monotonic
function of time and temperature. The temperature only plays the role of slowing down
dynamics, therefore a change in temperature only changes the growth
law. Any time $t_w$ spent at a
temperature $T_1$ is equivalent to having spent an effective time
$t_{\rm eff}(t_w, T_1,T_2)$ at $T_2$ such that 
$\xi^2(t_{w},T_1)=\xi^2(t_{\rm eff},T_2)$.
Thus in temperature cycling protocols there is no trace of the chaotic effects
observed experimentally on the correlation length itself, see Fig. \ref{cicle}.

\subsection{Susceptibility}

In the previous section we have studied the correlation length $\xi(t_{w})$ that tells 
us how far the particle can go in a time $t_w$. This quantity is the analogue
of the size of the domains in a droplet coarsening description. 
However, in a.c. experiments after a negative temperature shift one probably 
observes how the `domain walls' reconform on a scale which is small  compared 
to $\xi(t_{w})$. The susceptibility defined in \eq{eq24} probes these `small' 
length scales and might show an interesting behaviour during temperature cycling, 
not revealed by $\xi(t_{w})$ (see Fig \ref{cicle}). The study of the `response' function 
\eq{eq24} is useful because the results admit an intuitive interpretation in terms of 
the evolution of ${\cal P}(x,t_w)$, which is the quantity that keeps track 
of the thermal history of the system. 
 
In order to compare with experiments, one should be in the following 
conditions: long waiting times $t_w$ and low frequencies $\omega$ 
(as compared to microscopic timescales), but such that $\omega\;t_w\;\gg\;1$. This last
condition is imposed by the fact that a harmonic response can only be
measured on a time larger than one oscillation period. This also ensures that
one is in a regime where the violations of the Fluctuation Dissipation Theorem are weak
and one can identify the fluctuation that we measure, \eq{eq24}, to a {\it response} \cite{VH96}.

\begin{figure}
\begin{center}

\rotatebox{270}{
\includegraphics*[scale=0.5]{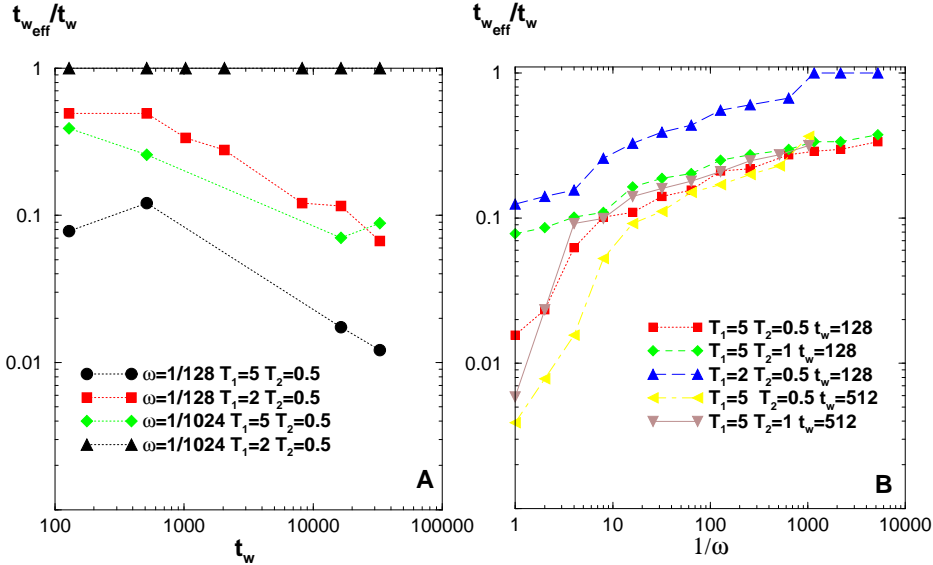}}
\caption{  Normalized effective times $t_{\rm eff}/t_{w1}$: plot A) at fixed $\omega$ versus 
$t_{w1}$, plot B) at fixed $t_{w1}$ versus $1/\omega$.   
\label{teffn}}
\end{center}
\end{figure}

 From the results of the
 simulations we observe that the effect of aging at temperature $T_1$
 on the relaxation at $T_2$ depends strongly on the temperature
 difference and on the waiting time. This effect can be quantified by
 defining an effective time. For instance, when cooling the system
 from $T_1$ to $T_2$ one expects that if the system is completely
 rejuvenated, the relaxation curve $\chi_{T_1,T_2}(\omega,t_{w2})$ should
 correspond to that obtained after quenching from high temperature
 $\chi_{\infty,T_2}(\omega,t_{w2})$. (Here $t_{w2}$ is counted from the time at
 which the system reaches $T_2$). However, if the relaxation at $T_1$
 affects aging at $T_2$ then rejuvenation is only partial and the new
 relaxation corresponds to that of the system
 after aging during an effective time $t_{\rm eff}$ at $T_2$,
 $\chi_{\infty,T_2}(\omega,t_{w2}+t_{\rm eff})$. Thus if $t_{\rm eff}=0$, 
 rejuvenation is complete. In Figure \ref{teffn} we
 show how this effective time is measured. Note that only the late part of the curves can 
 be superimposed: there is a transient 
that cannot be accounted for using an effective time. A similar effect can be observed 
in spin-glasses. The same effective time can also be defined when heating back the system, as
a measure of memory recovery.

\begin{figure}
\begin{center}
\rotatebox{270}{
\includegraphics*[scale=0.4]{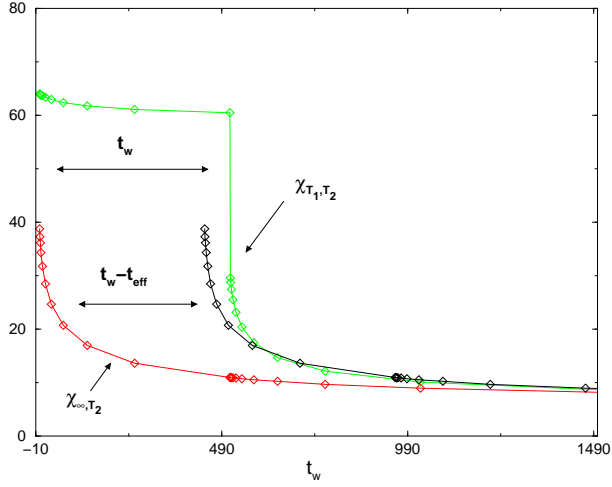}}

\caption{ Measure of the effective time when cooling the system 
down to $T_2=0.5$ after having spent $t_{w1}=512$ at $T_1=5$ at $\omega=1/128$.
Note that only the late part of the curves can be superimposed: there is a transient 
that cannot be accounted for using an effective time.
\label{teff}}
\end{center}
\end{figure}

Experimentally, the observed facts in spin glasses are the 
following. 

\begin{itemize}  

\item When cooling the sample from $T_1<T_c$ to $T_2<T_1$, $t_{\rm eff}$ is
larger than $t_{w1}$ when $\Delta T$ is small enough, indicating that aging at $T_1$ 
is more efficient because it is faster at higher temperature \cite{Sasaki02}. However,
as $\Delta T$ increases, $t_{\rm eff}$ starts decreasing towards zero. In other words, for 
large enough $\Delta T$, {\em rejuvenation} is complete. 
A related phenomenon is the absence of cooling rate effects. The relaxation at a low temperature does not depend on the cooling history. Only the very last steps at nearby temperatures determine the final relaxation \cite{VH96}. 

\item When heating back from $T_2$ to $T_1$ the system resumes its
relaxation, for large enough $\Delta T$, as if the stay at lower temperatures
did not take place (perfect {\it memory}). As $\Delta T$ is reduced, an 
effective age $< t_{w2}$ that accounts for the time spent at $T_2$ must 
be added. Furthermore, a small transient appears at short times, that 
was called `memory anomaly' in \cite{Sasaki02}. Surprisingly, this anomaly 
is non monotonous with $\Delta T$: for small $\Delta T$, the reference 
curve corresponding to perfect memory is reached from {\it below}, whereas for
larger $\Delta T$, it is reached from {\it above} \cite{Sasaki02}.
Finally, for still larger $\Delta T$'s, memory is perfect, as stated above. 

Note that if the time $t_{w1}$ spent at $T_1$ in the first stage of the cycle is very small, one expects, from the previous discussion, to see `rejuvenation' after a 
positive temperature shift, since the system has kept the memory of the age it
had on the way down, i.e. $t_{w1} \ll t_{w2}$. This rejuvenation after a positive 
$\Delta T$ has been observed many times experimentally.

\end{itemize}

Let us now turn to the results of the simulations, following the above presentation
of experimental data.

\begin{itemize}

\item The data of Fig \ref{dtn} corresponds to temperature cycles with 
$T_1=5$, $T_2=0.5$ or $T_1=2$, $T_2=0.5$. The main
observation here is that clear rejuvenation is observed, with an effective shift 
time that decreases as $\Delta T$ increases, as in the experiments. Intuitively,
this corresponds to the fact that since the potential is self-similar, the
{\it local} dynamics probed by $\chi(\omega,t_w)$ is not sensitive to 
the depth of the potential valley that is currently occupied. Therefore, aging 
at $T_1$ has already selected some low-lying valleys, but the intra-valley 
dynamics is insensitive to this. This argument for rejuvenation based on
a hierarchical energy landscape, and 
in the absence of temperature chaos has been put forward in \cite{Vincent,BD95,B99,BD01},
and has been confirmed numerically in the multi-level trap model in \cite{SN00}. 
Lower frequencies, that correspond to larger length scales, are less 
easily rejuvenated, as expected, since the separation of time scales is not as sharp:
see Figure \ref{teffn}, where we show the normalized effective waiting times as a function of 
both $t_{w1}$ and ${1}/{\omega}$. 

\item When heating back the system to the initial temperature $T_1$, some memory
is observed. However (i) when the temperature difference is not very large, some
effective time, accounting for the period spent at $T_2$, must be included, as in the
experiments, and (ii) a strong transient `memory anomaly'
is observed, even for quite large $\Delta T$'s (see plots A and B in figure \ref{dtn}). 
Note that we have always observed 
this memory anomaly to be negative, i.e; the reference curve is reached from
below. This memory anomaly is defined as:
\begin{equation}
\Delta \chi=\chi(t_{w1}+t_{w2}+1/\omega)-\chi(t_{w1}^-)~~,
\label{dchi}
\end{equation}
where $\chi(t_w={t_{w1}^-})$ is the susceptibility just before the quench and $\chi(t_{w1}+t_{w2}+1/\omega)$ corresponds to the first possible measurement at frequency $\omega$ after heating back to $T_1$.
In Figs. \ref{deltachi} and \ref{deltachir} the dependence of the memory anomaly on different parameters, 
including the frequency, is shown.

In fig. \ref{deltachi}  plot we show how the memory anomaly varies with the time $t_{w2}$ spent at $T_2$,
for a fixed $\omega=1/128$ and $t_{w1}=1024$, for the largest $\Delta T$. From the inset, we see
that the larger $\Delta T$, the smaller $|\Delta \chi|$, which agrees with the interpretation 
that the strong rejuvenation effect found for large $\Delta T$ arises from the separation 
of length scales. On  the plot in fig. \ref{deltachir} we show how the memory anomaly depends on the frequency $\omega$. Since the susceptibility at $t_{w1}$ itself depends on $\omega$ we plot the relative variation of the susceptibility  with respect to $\chi(t_{w1}^-)$ at fixed $t_{w1}=t_{w2}=1024$. In the inset we show $\Delta \chi/\chi(t_{w1}^-)$ versus $\omega$. Note that $\Delta \chi/\chi(t_{w1})$ is always 
negative and decreases with increasing $1/\omega$. In \cite{Sasaki02}, Sasaki et al. find that in
the anomaly can be both positive and negative when one works in the vicinity of a 
transition temperature. In the Sinai model we have not been able to 
observe such a positive anomaly. For smaller $\Delta T$'s these effects are blurred because 
length scale separation becomes weak. Aging at different temperatures is cumulative and
$\Delta \chi/\chi$ is also larger (see fig \ref{dtn} for the $T_1=2\to T_2=0.5$ cycle). For these
smaller $\Delta T$'s, we have found that the memory anomaly becomes non monotonous with frequency. 

\begin{figure}
\begin{center}
\rotatebox{270}{
\includegraphics*[scale=0.5]{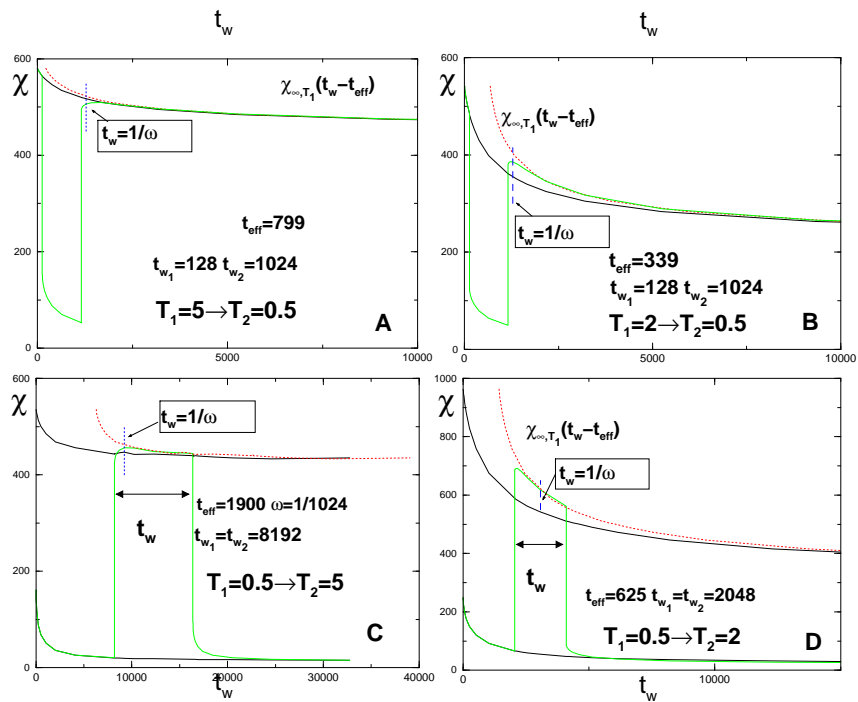}
}
\caption{Negative and positive temperature cycles.  Plots A and 
B cycles with $\Delta T<0$, data for $\omega=1/128$, 
$t_{w_1}=128, t_{w_2}=1024$ and for cycle A: $T_1=5$, $T_2=0.5$
and B: $T_1=2$, $T_2=0.5$.  Plots C and D cycles with $\Delta T>0$, 
for cycle C:  $T_1=0.5$, $T_2=5$, $\omega=1/1024$ 
and $t_{w_1}=t_{w2}=8192$ and D: $T_1=0.5$, $T_2=5$, $\omega=1/1024$ 
and $t_{w_1}=t_{w2}=8192$
\label{dtn}}
\end{center}
\end{figure}
\begin{figure}
\begin{center}
\rotatebox{270}{
\includegraphics*[scale=0.4]{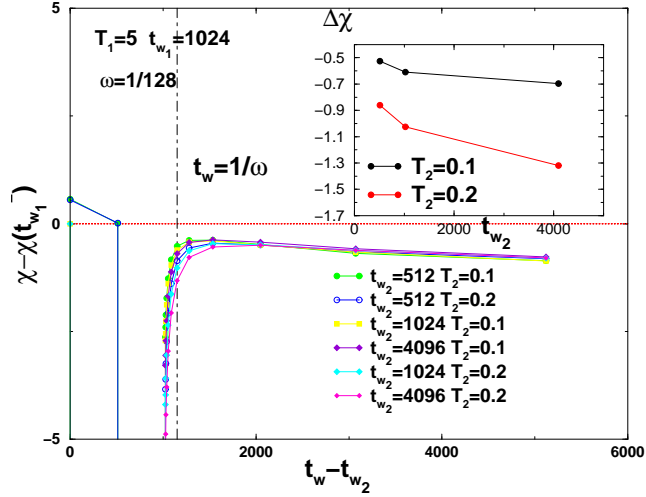}
}
\caption{Memory anomaly in a negative 
temperature cycle at fixed $\omega$. We plot the $\chi-\chi(t_{w1}^-)$ at fixed $\omega=1/128$ and 
$t_w1=1024$ 
for the temperature changes $T_1=5$ $T_2=0.1,0.2$ for different $t_{w2}=512, 1024,4096$. 
In the inset we plot $\Delta \chi$ as defined in \eq{dchi} for the different $t_{w2}$. 
\label{deltachi}}
\end{center}
\end{figure}
\begin{figure}
\begin{center}
\rotatebox{270}{
\includegraphics*[scale=0.4]{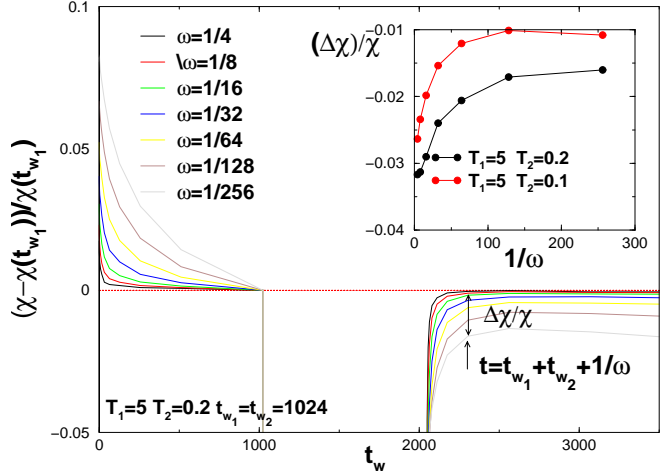}
}
\caption{Memory anomaly in a negative 
temperature cycle at fixed $\Delta T$.  $\Delta \chi/\chi(t_{w1}^-)$  for the cycle $T_1=5$ $T_2=0.2$ with $t_{w1}=t_{w2}=1024$ we plot $\Delta \chi/\chi(t_{w1}^-)$ for different frequencies, $\omega=1/4 ....1/256$. In the inset we plot $\Delta\chi/\chi$ 
versus $1/\omega$ for this case and for the cycle  $T_1=5$ $T_2=0.1$ with the same parameters.
\label{deltachir}}
\end{center}
\end{figure}

\item In the positive temperature cycle experiments 

We have also observed that the third stage is completely independent of $t_{w1}$.
Heating back the system to $T_2$ erases the initial aging accumulated at $T_1$.
This is expected, since the dynamics at $T_2$ allows the system to leave the traps 
that it had slowly explored at $T_1$. The new relaxation at $T_1$ is aged, 
but the age is only due to the effect of aging at $T_2$. This is similar to
the effect observed experimentally. 
  
\end{itemize}

\section{Discussion}

In this paper we studied in detail the role of temperature shifts in the simplest 
model where the energy landscape is explicitely hierarchical, namely the Sinai
model. This model has both attractive features (there are valleys within valleys 
in a strict self-similar sense), but also one important drawback: there is no
phase transition so that the model, in the large-size limit, 
is effectively at zero temperature. Therefore, in this limit, temperature shifts 
do not lead to interesting phenomena: entropy in this model does not play any role, 
so that excitations have an enormous free-energy cost and cannot be favored  by the 
temperature perturbation, contrarily to what happens in a closely related model,  the 
Directed Polymer \cite{SY02} (see also \cite{SM02}).

Nevertheless, for finite sizes/finite times, some interesting crossover phenomena
qualitatively reproduce the spin-glass phenomenology. In particular, dynamical 
rejuvenation effects in the absence of temperature chaos are observed. This rejuvenation
is ascribed to the local dynamics, which is insensitive (due to the self similar nature
of the potential) to the particular valley that has been reached during aging at a
higher temperature. Still, the separation of time scales/length scales with 
temperatures is much weaker that in experimental spin-glasses \cite{BD01}, 
partly due to the rather modest time scales investigated in the present study. 
Correspondingly, abrupt rejuvenation as the temperature is decreased and strict
memory when the temperature is cycled cannot be achieved. Rejuvenation and 
memory are present {\it in embryo}. 

In a word, the Sinai model is a smeared out version of the multi-level trap 
model studied in \cite{BD95,SN00}, where the sequence of critical temperatures
is replaced by a gradual freezing in the details of the fractal landscape. 
One could have studied the effect of temperature shifts in a one dimensional
potential with logarithmic correlations (rather than linear, as in the Sinai case).
This model was studied in details in \cite{CL01} and was shown to exhibit a true
transition temperature. We expect the resulting temperature shift 
phenomenology to be very close to that of the Random Energy Model, explored in \cite{SB01}. 
An extension of this model, that has an 
infinite sequence of phase transitions, will be presented in the near future \cite{B03}.

\vskip 5pt

{\bf Acknowledgments}

We thank L. Berthier, V. Dupuis, O. Martin, M. Sasaki, E. Vincent and H. Yoshino
for discussions. H. Yoshino participated to an early stage of the project.
M. S. thanks the MEC for grant AP98-36523875. F. R wishes to thank the
Ministerio de Ciencia y Tecnolog\'{\i}a in Spain, project BFM2001-3525
and Generalitat de Catalunya for financial support.

\setcounter{equation}{0}
\renewcommand{\theequation}{A-\arabic{equation}}
\section*{Appendix A: Calculation of the Entropy Fluctuations in the deepest well. } 

In this appendix we compute the average entropy fluctuations of 
the Sinai model in the large $L$ limit. As we have already said in section 
\ref{fe} these entropy fluctuations are constant in this limit. 
For this computation we only need to consider the contribution to the 
partition function arising from the deepest minima. In a recent paper \cite{MD02}  
Monthus and Le Doussal compute the probability distribution of the partition function 
of the deepest well considering that the origin in energies is at the bottom of the potential. 
They compute  $P({\cal Z})$ obtaining that the Laplace transform is 
\begin{equation}\int_0^\infty d{\cal Z} e^{-t {\cal Z}}\;P({\cal Z})=
\frac{1}{I_0^2\left(a\;\sqrt{t}\right)}~~~~a=\frac{2}{\beta\sigma}~~,
\label{a1}
\end{equation}
where $\beta=1/T$ and $\sigma^2$ is the variance of the random force. 
This result implies that we can work in terms of a dimensionless partition function  
$z={\cal Z}/{a^2}$ to obtain,
\begin{equation}\int_0^\infty dz e^{-t z}p(z)=\frac{1}{I_0^2\left(\sqrt{t}\right)}~~~
\label{a2}~~.
\end{equation}
Since the energy of the absolute minimum is set to zero, the entropy fluctuations can be 
directly computed as,
\begin{equation}\overline{ S(T)^2}-\left(\overline{S(T)}^2\right)=\overline{\ln^2 {\cal Z}}-
\overline{\ln {\cal Z}}^2~~.
\label{a3}
\end{equation}
Now in terms of the dimensionless partition function we have that $\ln {\cal Z}=
\ln z+ \ln a^2$, thus when computing fluctuations of averages over $z$ terms which 
depend on $a$ (and thus on $T$) cancel to yield,
\begin{equation}\overline{ S(T)^2}-\left(\overline{S(T)}^2\right)=\overline{\ln^2 z}-\overline{\ln z}^2=
\int_0^\infty dz \;p(z)\;\left(\ln z\right)^2-
\left(\int_0^\infty dz \;p(z)\;\ln z\right)^2~~~~,
\label{a4}
\end{equation}
which is a constant as we expected.
This constant can be evaluated by computing the averages of $\ln z$ and $\ln^2 z$. 
The starting point of the calculation is Derrida's integral representation of the logarithm,
\begin{equation}\ln {\cal Z}=\lim_{q\to 0}\int_q^\infty dt\;\frac{1}{t}\left(e^{-t}-e^{-t {\cal Z}}\right)~~,
\label{a5} 
\end{equation}
When we perform the average of $e^{-t z}$ we obtain the Laplace Transform given in 
\eq{a2} thus
\begin{equation} 
\overline{\ln z}=\lim_{q\to 0}\int_q^\infty dt\;\frac{1}{t}\left(\frac{1}{I_0^2
\left(\sqrt{t}\right)}-e^{-t}\right).
\label{a6}
\end{equation}
This integral can be split into two parts, the exponential integral
\begin{equation}\int_q^\infty dt\;\frac{1}{t}e^{-t}=-\gamma -\ln q+q+\frac{q^2}{2} +...~~~ \lim\; q\to 0 ~~,
\label{a7}
\end{equation}

where $\gamma$ is the Euler constant, and the part with the Bessel Function,  
\begin{eqnarray}
\int_q^\infty dt\;\frac{1}{t}\frac{1}{I_0^2\left(\sqrt{t}\right)}=2\;\
\int_{ \sqrt{q}}^\infty dt\;\frac{1}{t}\frac{1}{I_0^2\left(t\right)}
=2\;\int_{\sqrt{q}}^\infty dt\; \left(-\frac{ d}{dt}\frac{K_0(t)}{I_0(t)}\right)=
2\;\frac{K_0(\sqrt{q})}{I_0(\sqrt{q})}\nonumber\\
= -\ln q+2\ln 2-2\gamma+(\psi(2)-1)\frac{q}{8}+...
\label{a8}
\end{eqnarray}
where we have used the Wronskian property of the Bessel functions
\begin{equation}-\frac{1}{t}=K'_{\nu}(t)\;I_{\nu}(t)-K_{\nu}(t)\;I'_{\nu}(t)~~.
\label{a9}
\end{equation}
Adding this two contributions we obtain that the average entropy reads,
\begin{equation}\overline{S(T)}=\overline{\ln {\cal Z}}=-2\ln a - 2\ln 2+\gamma~~.
\label{a10}
\end{equation}
To compute the average $\ln^2z$ we have to evaluate the following integral,
\begin{equation}\overline{\ln^2 z}=\lim_{q\to 0,\;p\to 0} \int_q^\infty dt\int_p^\infty dt'
\frac{1}{t\;t'} \left(e^{-t-t'}-\frac{e^{-t'}}{I_0^2(\sqrt{t})}-
\frac{e^{-t}}{I_0^2(\sqrt{t'})}+\frac{1}{I_0^2(\sqrt{t+t'})}\right)~~.
 \label{a11}
\end{equation}
The contribution arising from the first three terms in the integral can be easily 
evaluated from expressions \eq{a6} and \eq{a7} to obtain:
\begin{eqnarray}
\int_q^\infty dt\int_p^\infty dt'\frac{e^{-t-t'}}{t\;t'}&=&\gamma^2+\gamma{\ln q p}+
\ln q\;\ln p+{\cal O}(q,p)\nonumber\\
\int_q^\infty dt\int_p^\infty dt'\frac{e^{-t}}{t\;t'}\frac{1}{I_0^2(\sqrt{t'})}&=&
(\gamma+\ln q)(2\gamma-2\ln 2+\ln p)\nonumber\\
\int_q^\infty dt\int_p^\infty dt'\frac{e^{-t'}}{t\;t'}\frac{1}{I_0^2(\sqrt{t})}&=&
(\gamma+\ln p)(2\gamma-2\ln 2+\ln q)  ~~. 
\label{a12}
\end{eqnarray}
The last term is somewhat more tricky and can be evaluated as follows,
\begin{eqnarray}
\int_q^\infty dt\int_p^\infty dt'\frac{1}{I_0^2(\sqrt{t+t'})}=2\;\int_q^\infty dt
\frac{1}{t}\frac{K_0(\sqrt{t+p})}{I_0(\sqrt{t+p})}+2\;\int_p^\infty dt'\frac{1}{t'}
\frac{K_0(\sqrt{t'+q})}{I_0(\sqrt{t'+q})}~~~.
\label{a13}
\end{eqnarray}
In the limit $p\to 0$ the integral can be Taylor expanded yielding,
\begin{equation}\int_q^\infty dt\frac{1}{t}\frac{K_0(\sqrt{t+p})}{I_0(\sqrt{t+p})}=\int_q^\infty dt
\frac{1}{t}\frac{K_0(\sqrt{t})}{I_0(\sqrt{t})}+{\cal O}(p) 
\label{a14}
\end{equation}
where the terms of order $p$ do not contribute when we set $p$ to $0$. Hence we can 
evaluate expression \eq{a13} at $p=0$ to obtain,
\begin{equation}\int_q^\infty dt\frac{1}{t}\frac{K_0(\sqrt{t})}{I_0(\sqrt{t})}=\frac{1}{2}\ln^2 q-2\;
\ln 2 \ln q+2\;\gamma\;\ln q+2\;\zeta~~~~~~\zeta=0.2415...
\label{a15}
\end{equation}
Adding all the contributions and taking the double limit $q\to 0$ and $p\to 0$ we obtain 
the following result for $\overline{\ln^2 z}$,
\begin{equation} \overline{\ln^2z}=4\;\zeta -3\;\gamma +4\;\gamma\;\ln 2~~,
\end{equation}
which finally yields the entropy fluctuation,
\begin{equation}K=\overline{S^2}-\overline{S}^2=4\;\left(\zeta-\left(\gamma-\ln2\right)^2\right)=0.912784....
\end{equation}

\setcounter{equation}{0}
\renewcommand{\theequation}{B-\arabic{equation}}
\section*{Appendix B: High temperature expansion of the distance shift} 
\label{apb}
We want to compute the high temperature behavior of the average distance shift between systems at two different temperatures $d_{1,2}$ as defined in eq.\eq{eq2},
\begin{equation} 
\overline{(\langle x\rangle_1-\langle x\rangle_2)^2}
\label{def}
\end{equation}
In the high temperature limit $\beta\to 0$ a simple Taylor expansion yields,
\begin{equation} 
\langle x\rangle=\frac{1}{L}\left(\sum_i x_i-\beta \left(\sum x_i V_i
-\sum x_i\sum \frac{V_i}{L}\right)\right)
\end{equation} 
where $V_i$ is the random potential at position $x_i=i/L$.\\
Therefore for \eq{def} we obtain, 
\begin{eqnarray} 
\overline{(\langle x\rangle_1-\langle x\rangle_2)^2}&=&\Delta \beta^2 \frac{1}{L^2}\overline{\left(\sum_i x_i V_i
-\sum_i x_i\sum \frac{V_i}{L}\right)^2}\nonumber\\
&=& (\Delta \beta)^2 \frac{\sigma^2}{120} \frac{(L^2-1)(L^2+10 L+11))}{L^3}  ~~,          \end{eqnarray}
where in the last equality we have used the relation $\overline{V_i\;V_j}=\sigma^2\;\min\{i,j\}$ which holds for Gaussian distributed forces with zero mean. For large $L$ we recover expression \eq{eq6},
\begin{equation}
\overline{(\langle x\rangle_1-\langle x\rangle_2)^2}=\frac{\sigma^2 }{120}(\Delta \beta)^2 \;L
\end{equation}


\begin{thebibliography}{99}

\bibitem{Gold69} M. Goldstein, J. Chem. Phys. {\bf 51}, 3728 (1969)
\bibitem{SW82} F. H. Stillinger, T. A. Weber, Phys. Rev.  A {\bf 25}, 978 (1982)
\bibitem{KL96} J. Kurchan, L. Laloux, J. Phys. A: Math. Gen. {\bf 29}, 1929 (1996)
\bibitem{I01} A. Cavagna, Europhys. Lett. {\bf 53}, 490 (2001), T.S. Grigera, A. Cavagna, I. Giardina,
G. Parisi, Phys. Rev. Lett, {\bf 88}, 055502 (2002) and refs. therein.
\bibitem{MPV} M. M\'ezard, G. Parisi and M. A. Virasoro, {\em Spin Glass Theory and Beyond}, World-Scientific, (Singapore 1987).
\bibitem{Vincent} Ph. Refregier, E. Vincent, J. Hammann and M. Ocio,
 J. Phys. France {\bf 48}, 1533 (1987); E. Vincent, J.-P.
Bouchaud, J. Hammann and F. Lefloch,  Phil. Mag. B {\bf 71},
489 (1995).
\bibitem{BD95} J.-P. Bouchaud and D. S. Dean, J. Phys. I (France)
{\bf 5}, 265 (1995).
\bibitem{CK99}  L. Cugliandolo, J. Kurchan, Phys. Rev. {\bf B 60}, 922 (1999).


\bibitem{HV00}J. Hammann, E. Vincent, V. Dupuis, M. Alba, M. Ocio and J.-P. Bouchaud, J. Phys. Jpn. {\bf 69}, 206 (2000) Suppl. A.
\bibitem{JV98}K. Jonson, E. Vincent, J. Hammann, J.-P. Bouchaud, and P. Norblad, Phys. Rev. Lett. {\bf 81}, 3243 (1998).
\bibitem{VH96} E. Vincent, J. Hammann, M. Ocio, J.-P. Bouchaud and L. F. Cugliandolo in {\em Complex Behaviour of Glassy Systems} eds.: M. Rub\'{\i} and C. J. P\'erez, (Lecture Notes in Physics, Springer-Verlag, 1996).
\bibitem{Parker} A. Parker, V. Normand, in preparation.
\bibitem{FH88} D. S. Fisher and D. A. Huse, Phys. Rev. B {\bf 38}, 386 (1988); D. S. Fisher and D. A. Huse, Phys. Rev. B {\bf 38}, 373 (1988). 
\bibitem{BM87} A. J. Bray and M. A. Moore, Phys. Rev. Lett. {\bf 58}, 57 (1987).
\bibitem{B99} J.-P. Bouchaud, in {\em Soft and Fragile Matter}, ed.: M. E. Cates and M.R. Evans (Institut of Physics Publishing, Bristol and Philadelphia, 2000). 
\bibitem{BD01} J.-P. Bouchaud, V. Dupuis, J. Hammann, E. Vincent, Phys. Rev. B {\bf 65}, 024439 (2002) (and e-print cond-mat/0106539). 
 \bibitem{YL01} H. Yoshino, A. Lemaitre , J.-P. Bouchaud, Eur. Phys. J. B  {\bf 20}, 367-395 (2001).

\bibitem{BB02} L. Berthier, J.-P. Bouchaud, cond-mat/0202069
\bibitem{BH02} L. Berthier and P. C. W. Holdsworth, Europhys. Lett. {\bf 58} 35 (2002).  
\bibitem{SB01} M. Sales and J.-P. Bouchaud, Europhys. Lett. {\bf 56}, 181 (2001). 
\bibitem{SN00} M. Sasaki and K. Nemoto, t J. Proc. Soc. Jpn {\bf 69}, 2283 (2000)
\bibitem{K01} M. Kawasaki , J. Phys. Soc. Jpn.  {\bf 70}, 1762-1767 (2001).
\bibitem{FH91} D. S. Fisher and D. A. Huse, Phys. Rev. B {\bf 43}, 10728 (1991).
\bibitem{SY02} M. Sales and H. Yoshino, Phys. Rev. E {\bf 65}, 066131 (2002).
(also cond-mat/0203371).
\bibitem{dSB02} R. da Silveira, J.-P. Bouchaud, still in preparation.
\bibitem{KONDOR89} I. Kondor, J. Phys. A: Math. Gen. {\bf 22}, L163 (1989).
\bibitem{R01} T. Rizzo, J. Phys. A: Math. Gen. {\bf 34},  5531-5549 (2001).
\bibitem{RITORT94} F. Ritort, Phys. Rev. B {\bf 50}, 6844 (1994).
\bibitem{NY97} M. Ney-Nifle and A. P. Young, J. Phys. A: Math. Gen. {\bf 30}, 5311 (1997) 
\bibitem{NIFLE98} M. Ney-Nifle, Phys. Rev. B {\bf 57}, 492 (1998).  
\bibitem{BM00} A. Billoire, E. Marinari,  J. Phys. A: Math. Gen. {\bf 33}, L265 (2000),
see also cond-mat/0202473.
\bibitem{SM02} M. Sasaki, O. Martin, e-print cond-mat/0206316
\bibitem{YJ02}  P. E. J\"onsson, H. Yoshino, P. Nordblad, preprint cond-mat/0203444 (accepted for publication in Phys. Rev. Lett 2002). 
\bibitem{PR01} M. Picco, F. Ricci-Tersenghi and  F. Ritort, Phys. Rev. B {\bf63 }, 174412 (2001); Eur. Phys. J. B {\bf 21}, 211 (2001). 
\bibitem{BC90} J.-P. Bouchaud, A. Comtet, A. Georges and P. le Doussal, Ann. Phys. {\bf 201}, 285 (1990).
\bibitem{BG90} J.-P. Bouchaud, A. Georges, Phys. Rep. {\bf 195}, 127 (1990) 
\bibitem{LD98} L. Laloux  and P. Le Doussal, Phys. Rev. E {\bf 57}, 6296 (1998). 
\bibitem{DM99} P. Le Doussal , C. Monthus and D. Fisher, Phys. Rev E {\bf 59}, 4795 (1999).
\bibitem{MD02} C. Monthus, P. Le Doussal, cond-mat/0202295.
\bibitem{MP91} M. M\'ezard, G. Parisi, J. Physique I {\bf 1}, 809 (1991)
\bibitem{ClD96}  L. F. Cugliandolo and  P. Le Doussal,
 Phys. Rev. E {\bf 53}, 1525  (1996).
\bibitem{CC02} F. Corberi, C. Castellano, E. Lippiello and  M. Zannetti, e-print cond-mat/0203250 and references therein.
\bibitem{Lud01} S. Luding, M. Nicolas, O. Pouliquen, cond-mat/0003172. 
\bibitem{LN01} D. Lubensky, D. R. Nelson, e-print cond-mat/0107423

\bibitem{Mezard} M. M\'ezard, J. Phys. France {\bf 51}, 1831 (1990)
\bibitem{Parisi} G. Parisi, J. Phys. France {\bf 51}, 1595 (1990) 
\bibitem{BO} J.-P. Bouchaud, H. Orland, J. Stat. Phys. {\bf 61}, 877 (1990)
\bibitem{OM93} G. Oshanin, A. Mogutov and M. Moreau, J. Stat. Phys. {\bf 73}, 379 (1993).
\bibitem{CM98} A. Comtet, C.  Monthus, M. Yor,  J. of App. Prob. {\bf 35}, (1998) 255-271.
\bibitem{BK} K. Broderix, R. Kree, cond-mat/9507099
\bibitem{hajime} T. Komori, H. Yoshino and H. Takayama,
J. Phys. Soc. Jpn. {\bf 68}, 3387 (1999); {\bf 69}, 1192 (2000);
{\bf 69}, Suppl. A 228 (2000).
\bibitem{D01} V. Dupuis, E. Vincent, J.-P. Bouchaud, J. Hammann,
A. Ito and H. A. Katori, Phys. Rev. B {\bf 64}, 174204 (2001).
\bibitem{YN01} P. E. J\"onsson, H. Yoshino, P. Nordblad, 
H. Aruga Katori, and A. Ito, Phys. Revl. Lett. {\bf 88}, 257204 (2002) (and preprint cond-mat/0112389).
\bibitem{Feller} W. Feller  in {\em An Introduction to Probability Theory and Its Applications}, New York: Wiley, (1968). 
\bibitem{Sasaki02} M. Sasaki, V. Dupuis, J.-P. Bouchaud, E. Vincent, 
e-print cond-mat/0205137
\bibitem{CL01} D. Carpentier, P. Le Doussal,  Phys. Rev. E {\bf 63}, 026110 (2001)
\bibitem{B03} J.-P. Bouchaud, in preparation

\end{thebibliography}
\end{document}